\documentclass{article}
\usepackage{amssymb}
\usepackage{amsmath}

\input{tcilatex}

\begin{document}

\title{Peculiar properties of the Josephson junction at the transition from
0 to $\pi $ state}
\author{A. Buzdin \\
Institut Universitaire de France and \\
Universit\'{e} Bordeaux I, CPMOH, UMR 5798, \\
33405 Talence, France}
\maketitle

\begin{abstract}
It is demonstrated that in the diffusive
superconductor-ferromagnet-superconductor (S/F/S) junctions the
current-phase relation is practically sinusoidal everywhere except in a
narrow region near the $0-\pi $ transition. In this region the second
harmonic dominates the scenario of the $0-\pi $ transition. We predict a
first order transition for the S/F/S junctions with a homogeneous F barrier.
However, in real junctions a  small modulation of the thickness of the
barrier may favor the continious $0-\pi $ transition and the realisation of
the Josephson junction with an arbitrary ground state phase difference.

The performed calculations of the second harmonic amplitude provide a
natural explanation of the recent contradictory results on the second
harmonic measurements.
\end{abstract}

In usual Josephson junctions (JJs) at equilibrium the phase difference of
the superconducting order parameter on the two banks is zero \cite%
{JJTextbook}. However, the situation may be drastically different for JJs
with a ferromagnetic interlayer (S/F/S junctions), where for some intervals
of the exchange field $h$ and F layer thickness $d,$ the ground state
corresponds to the phase difference equal to $\pi $ (\textquotedblright $\pi 
$-junctions\textquotedblright ) \cite{SFSpi-junction},\cite{Buz-Kupr}. This
phenomenon is related to the damping oscillatory behavior of the Cooper pair
wave function in a ferromagnet (for more references and as reviews see \cite%
{RevBuz},\cite{RevGol}). The first experimental evidence of a $0-$ $\pi $
transition in S/F/S ($Nb-Cu_{x}Ni_{1-x}-Nb$) JJs was obtained by Ryazanov 
\emph{et al.} \cite{ryazanov} from the measurements of the temperature
dependence of the critical current. The $0-$ $\pi $ transition was signaled
by the vanishing of the critical current with the temperature decrease. Such
a behavior is observed for a F layer thickness $d$ close to some critical
value $d_{c}$. In fact, it simply means that the critical thickness $d_{c}$
slightly depends on the temperature. The temperature variation serves as a
fine tuning and permits to study this transition in detail. Recently,
thorough measurements of the critical current were performed in $%
Nb-Cu_{x}Ni_{1-x}-Nb$ junctions with $x=0.53$ and $d\sim 22nm$ \cite{Frolov}
and in similar junctions with smaller $x=0.48$ and $d\sim 17nm$ \cite%
{Sellier2} . The results were contradictory, since the critical current at $%
0-$ $\pi $ transition in \cite{Frolov} was zero, while in experiments \cite%
{Sellier2} a small critical current was observed.

In this Letter we elaborate a theory describing how the $0$ state is
transformed into the $\pi $ state. \textit{\ }It is demonstrated that the
critical current of the S/F/S JJs does not vanish at the transition and is
determined by the second harmonic term in the current-phase relation. This
second harmonic contribution decreases extremely strongly with the increase
of the thickness of \ the F layer and its exchange field. The corresponding
estimate for the critical current at the $0-$ $\pi $ transition in
experiments \cite{Frolov} \ gives the value well below the experimental
resolution. On the other hand, for the parameters of S/F/S junctions in Ref. 
\cite{Sellier2}, the calculated amplitude of the second harmonic is close to
the experimentally measured value. The $0-$ $\pi $ transition is
discontinuous for junctions with a homogeneous ferromagnetic barrier. In
real S/F/S junctions the modulation of the F layer thickness may provide a
contribution to the second harmonic with the opposite sign \cite{Buzdin
Koshelev},\cite{Mints1}. If this mechanism prevails then the $0-$ $\pi $
transition would be continuous. This means that by varying the JJ parameters
(e. g. temperature) it is possible to obtain the S/F/S junction with \textit{%
arbitrary ground state phase difference.} We also briefly discuss the
thermodynamics of the $0-$ $\pi $ transition. Note that previously the
current-phase relations in S/F/S junction were calculated numerically in 
\cite{Heikkila} and analytically for several special types of the composite
SF-FS junctions and short S/F/S junctions in \cite{RevGol},\cite{A. A.
Golubov}. However the theoretical approach to treat analytically the
diffusive S/F/S junction with the $0$ $-$ $\pi $ transition (most relevant
to the experiment) was lacking.

The current-phase relation for JJ is sinusoidal only near the critical
temperature $T_{c}$ \cite{JJTextbook} 
\begin{equation}
j(\varphi )=I_{1}\sin \varphi .  \label{sin}
\end{equation}%
At low temperature, the higher harmonic terms become more and more
important. The calculations of the current-phase relations of the S/F/S
junctions in a clean limit indeed reveal strongly non-sinusoidal $j(\varphi )
$ dependences at low temperatures \cite{SFSpi-junction},\cite{Radovic2001},%
\cite{Chtchelkatchev2001}. However, in the experiments \cite{Frolov},\ \cite%
{Sellier2} the ferromagnetic alloys are used as a F layer, and the dirty
limit is more appropriate for the description of this case. In such a limit
in a normal JJs, if the length $d$ of the weak link exceeds the
characteristic length $\xi _{1}$ of the decay of the Cooper pair wave
function, the critical current is small $j_{c}\sim \exp (-d/\xi _{1})$ and $%
j(\varphi )$ is practically sinusoidal \cite{RevGol}. We demonstrate that
the second harmonic contribution is very small $\sim \exp (-2d/\xi _{1})$.
Usually the role of the second harmonic is negligible and hardly observable.
However, in S/F/S\ junctions the first harmonic vanishes at the $0$ $-$ $\pi 
$ transition and the situation occurs to be very different - the
contribution of the second harmonic becomes predominant!

The general current-phase relation

\begin{equation}
j(\varphi )=I_{1}\sin \varphi +I_{2}\sin 2\varphi  \label{I1+I2}
\end{equation}
corresponds to the following phase dependent contribution to energy of the JJ

\begin{equation}
E_{J}(\varphi )=\frac{\Phi _{0}}{2\pi c}\left[ -I_{1}\cos \varphi -\frac{%
I_{2}}{2}\cos 2\varphi \right] .  \label{EJ}
\end{equation}
If we neglect the second harmonic term, then the $0$ state occurs for $%
I_{1}>0.$ Near a $0$ $-$ $\pi $ transition $I_{1}\rightarrow 0$ and the
second harmonics term becomes important. The critical current at the
transition $j_{c}=\left\vert I_{2}\right\vert $ and if $I_{2}>0$, the
minimum energy always occurs at \ $\varphi =0$ or $\varphi =\pi $ (Fig. 1).
In the opposite case ($I_{2}<0$) the transition from $0$ to $\pi $ state is
continuous and there is region where the equilibrium phase difference takes
any value $0<\varphi <\pi $. The characteristics of such a $"\varphi -$
junction$"$ are very peculiar \cite{Buzdin Koshelev}.

To describe the properties of the S/F/S\ junction in the diffusive limit we
use the Usadel equations\cite{Usadel}. Recent studies \cite{Sellier1}, \cite%
{Ryazanov2004} revealed a very strong variation of $j_{c}$ with the F layer
thickness, which implies strong magnetic scattering effects \cite%
{Ryazanov2005}. Assuming the presence of the relatively strong uniaxial
magnetic anisotropy in F layer we may neglect the magnetic scattering in the
plane perpendicular to the anisotropy axes (which mixes the spin up and down
Green functions) and the Usadel equation for the normal $G(x,\omega ,h)$ and
anomalous $F\left( x,\omega ,h\right) $ Green functions in the F layer is
(see for example \cite{RevBuz}):

\begin{eqnarray}
&&\left. -\frac{D_{f}}{2}\left[ G(x,\omega ,h)\frac{\partial ^{2}}{\partial
x^{2}}F\left( x,\omega ,h\right) -F(x,\omega ,h)\frac{\partial ^{2}}{%
\partial x^{2}}G\left( x,\omega ,h\right) \right] \right.   \notag \\
&&\left. +\left( \omega +ih+\frac{G(x,\omega ,h)}{\tau _{s}}\right)
F(x,\omega ,h)=0,\right.   \label{Usadel gen}
\end{eqnarray}%
where the $x$ axis is perpendicular to the junction plane and the F layer
corresponds to $-d/2<x<d/2$, $D_{f}$ is the diffusion constant in the F
layer and $\tau _{s}$ is the magnetic scattering time.\ In the spatially
uniform case, the equation (\ref{Usadel gen}) is equivalent to the one from
the Abrikosov-Gorkov theory \cite{A-G}. The equations (\ref{Usadel gen})
must be completed by the boundary conditions at the S/F interface \cite%
{Kup-Luk}. Below we consider two limiting cases: transparent interfaces and
large interface barriers. Moreover, assuming the normal state conductivity $%
\sigma _{f\text{ }}$ of the F-layer small compared to that of the S-layers, $%
\sigma _{s}$ $>>\sigma _{f\text{ }}$, we may neglect the influence of the
F-layer on S-layer, i.e. the Green functions in the left S-layer are $%
F_{s}\left( \omega \right) =\Delta e^{i\varphi /2}/\Omega $ , $G_{s}\left(
\omega \right) =sgn(\omega )\omega /\Omega $, where $\Omega =$ $\sqrt{\omega
^{2}+\left\vert \Delta \right\vert ^{2}}$ (for the right S-layer $\varphi
\rightarrow -\varphi $).

For transparent interfaces, the boundary conditions \cite{Kup-Luk} express
the continuity of the Green functions. At $T=T_{c}$ the equation for $%
F(x,\omega ,h)$ is linear and may be easily solved \cite{RevBuz}. Just below 
$T_{c}$ the nonlinear corrections in (\ref{Usadel gen}) are small and we may
apply the method similar to that, used in the problem of the non-linear
oscillator \cite{LanLifMech}. After some calculations we have:

\begin{equation}
F(x,\omega )=a\cosh (k_{1}x)+b\sinh (k_{2}x)+\frac{b^{2}-a^{2}}{32k^{2}}%
\left( \frac{1}{D_{f}\tau _{s}}+\frac{3}{2}k^{2}\right) \left( a\cosh
(3kx)+b\sinh (3kx)\right) ,  \label{Fnonl}
\end{equation}
where the complex wave-vectors $k_{1}$ and $k_{2}$ are determined by the
relations

$k_{1}^{2}-k^{2}=\frac{k^{2}\left( 5b^{2}-a^{2}\right) }{8}-\frac{\left(
3a^{2}+b^{2}\right) }{4D_{f}\tau _{s}}$, $k_{2}^{2}-k^{2}=\frac{k^{2}\left(
5a^{2}-b^{2}\right) }{8}-\frac{\left( 3b^{2}+a^{2}\right) }{4D_{f}\tau _{s}}$
. The non-linear effects make $k_{1}$ and $k_{2}$ different from the
wave-vector $k^{2}(\omega ,h)=2(\omega +ih+sgn(\omega )/\tau _{s})/D_{f}$ \
in the solution of the linear equation. The coefficients $a$ and $b$ are
determined from the boundaries conditions and the current-phase relation may
be then directly obtained from the formula for the supercurrent \cite{Usadel}%
.

In the limit $h>>T_{c}$ and in the absence of magnetic scattering the Cooper
pair wave function in ferromagnet decays and oscillates at the same
characteristic length \ $\xi _{f}=\sqrt{h/D_{f}}$ \cite{Buzdin}. The
magnetic scattering leads to the decrease of the decaying length $\xi
_{1}=\xi _{f}/\sqrt{\alpha +\sqrt{1+\alpha ^{2}}}$ and to an increase of the
oscillating length $\xi _{2}=\xi _{f}/\sqrt{\sqrt{1+\alpha ^{2}}-\alpha },$
where the dimensionless parameter $\alpha =1/(h\tau _{s})$. Note that the
product $\xi _{1}\xi _{2}=\xi _{f}^{2}$ is a constant and independent on $%
\alpha $. If the F layer thickness $d\gtrsim \xi _{1}$ the expression for
the first harmonic reads 
\begin{equation}
I_{1}=j_{0}\frac{\Delta ^{2}\left( 1+\alpha ^{2}\right) ^{1/4}}{2\sqrt{2}%
T_{c}^{2}}\exp (-d/\xi _{1})\left[ \sin \left( d/\xi _{2}+\Psi \right) -%
\frac{\alpha \Delta ^{2}}{96T^{2}\left( 1+\alpha ^{2}\right) ^{1/2}}\sin
\left( d/\xi _{2}-\Psi \right) \right] ,  \label{I1}
\end{equation}
where $j_{0}=4\pi T_{c}eSN(0)D_{f}/\xi _{f}$, the angle $\pi /4<\Psi <\pi /2$
is determined by the relation $\tan \Psi =\alpha +\sqrt{1+\alpha ^{2}}$, and 
$S$ is the area of the cross section of the junction and $N(0)$ is the
electron density of state per one spin projection. The amplitude of the
first harmonic (\ref{I1}) reveals an oscillatory decay as a function of the
F layer thickness. The sign change of $I_{1}$ signals the $0$ - $\pi $ \
transition. At $T\rightarrow T_{c}$ the critical thickness separating $0$
and $\pi $ phases is $d_{c}^{n}(T_{c})=\left( \pi n-\Psi \right) \xi _{2},$
which decrease with the decreases with temperature

\begin{equation}
d_{c}^{n}(T)=d_{c}^{n}(T_{c})-\frac{\Delta ^{2}\alpha \xi _{2}}{%
96T_{c}^{2}\left( 1+\alpha ^{2}\right) }.
\end{equation}
The second harmonic term is much smaller 
\begin{equation}
I_{2}=-j_{0}\frac{\Delta ^{4}}{96T_{c}^{4}}\exp (-2d/\xi _{1})\left[ \left(
d/\xi _{f}\right) \sin \left( 2d/\xi _{2}\right) +\frac{5\sin \left( 2d/\xi
_{2}+\Psi \right) +3\sin \left( 2d/\xi _{2}-3\Psi \right) }{4\sqrt{2\sin
\left( 2\Psi \right) }}\right] ,  \label{I2}
\end{equation}
and compared with $I_{1}$ it contains, in addition to the factor $\Delta
^{2}/T_{c}^{2}$, an exponentially small term $\exp (-d/\xi _{1})$. The
amplitude of $I_{2}(d)$ also reveals the oscillatory-like dependence similar
to $I_{1}(d)$ but with the decaying and oscillating lengths two time
smaller. It may be directly verified that at the $0$ - $\pi $ transition
(i.e. when $d=d_{c}^{n}$) $I_{2}$ is always \textit{positive}.

Now let us consider the limit of large S/F interface barriers which is
characterized by the parameter $\gamma _{B}$, related to the S/F boundary
resistance per unit area $R_{b}$ by $\gamma _{B}=R_{b}\sigma _{f}$. In this
case the boundary condition reads $\gamma _{B}\left( \partial F/\partial
x\right) _{x=d/2}$ $=F_{s}\left( x=d/2\right) G_{f}^{2}\left( x=d/2\right) $ 
\cite{Kup-Luk}. Performing the similar analysis as in the perfect
transparency case $(\gamma _{B}=0)$ we obtain in the limit $d\gtrsim \xi
_{1}:$

\begin{equation}
I_{1}=j_{0}\left( \frac{\xi _{f}}{2\gamma _{B}}\right) ^{2}\frac{\Delta ^{2}%
}{T_{c}^{2}\sqrt{2}\left( 1+\alpha ^{2}\right) ^{1/2}}\exp (-d/\xi _{1})\sin
\left( \Psi -d/\xi _{2}\right) .  \label{I1large G}
\end{equation}%
The second harmonic term at $\alpha \thicksim 1$ is of the order $%
I_{2}\thicksim j_{0}\left( \frac{\xi _{f}}{\gamma _{B}}\right) ^{4}\exp
(-2d/\xi _{1})$ and also oscillates (with a period $\pi \xi _{2}$, which is
two times smaller than that of $I_{1}$) and \ similar to the transparent
interface case is positive at the $0$ - $\pi $ transition.

For small $\alpha <<1$, the first $0$ - $\pi $ transition occurs at the F
layer thickness smaller than $\xi _{f}$ \cite{Buzdin}. In the considered
case (assuming $h,\tau _{s}^{-1}>>T$ ) the first $0$ - $\pi $ transition
occurs at $\ d=d_{c}^{0}=\xi _{f}\sqrt{3\alpha }$ and

\begin{equation}
I_{1}=j_{0}\left( \frac{\xi _{f}}{2\gamma _{B}}\right) ^{2}\frac{\Delta ^{2}%
}{2T_{c}^{2}}\frac{d_{c}^{0}-d}{\xi _{f}},  \label{10}
\end{equation}%
The second harmonic term at the $0$ - $\pi $ transition is also positive and 
$I_{2}\thicksim j_{0}\left( \frac{\xi _{f}}{\gamma _{B}}\right) ^{4}\frac{%
\Delta ^{4}}{T_{c}^{4}\sqrt{\alpha }}.$ The formula (\ref{I1large G},\ref{10}%
) are written for $T\lesssim T_{c}$ but the corresponding analysis is easily
generalized for all temperatures. Besides the change of numerical
coefficients the expressions for $I_{1}$ and $I_{2}$ remain the same.

Now we demonstrate how the obtained results permit to understand the
controversy in the experimental search of the second harmonic \cite{Frolov}, 
\cite{Sellier2}. From the thickness dependence of the critical current in
the series of $Nb-Cu_{0.47}Ni_{0.53}-Nb$ junctions \cite{Ryazanov2004},\cite%
{Ryazanov2005} we may estimate $\xi _{1}\thickapprox 1.4$ $nm$ and $\xi
_{2}\thickapprox 4.1$ $nm$. This gives $\xi _{f}\thickapprox 2.4$ $nm$, the
magnetic scattering parameter $\alpha \thickapprox 1.3$ and the exchange
field $h\thickapprox 600$ $K$. In the experiments \cite{Frolov} the
current-phase relation was measured near the second $0$ - $\pi $ transition
at $d\thickapprox 22$ $nm$ (the first transition occurs at $d\thickapprox 11$
$nm$ \cite{Ryazanov2005}). Therefore we may roughly estimate $%
I_{2}/I_{1}\sim 0.1\exp (-d/\xi _{1})\sim 10^{-8}$, which gives very small
value for $I_{2}\lesssim 10^{-11}$ $A$, well below the experimental
threshold. On the other hand the corresponding estimate for the first $0$ - $%
\pi $ transition at $d\thickapprox 11$ $nm$ is much more favorable for $%
I_{2} $ observation: $I_{2}/I_{1}\sim 10^{-4},$ and $I_{2}\sim 10^{-6}$ $A$.
Therefore it would be interesting to perform the similar measurements on the
junctions revealing temperature mediated first $0$ - $\pi $ transition.

In similar junctions, but with smaller $Ni$ concentration $x=0.48$ the
second harmonic at\ the $0$ - $\pi $ transition was reported for the F-layer
thickness $d\thickapprox 17$ $nm$ at $1.1$ $K$ \cite{Sellier2}. This is the
first (as a function of $d$) $0$ - $\pi $ transition. For junctions with $%
x=0.48$ \cite{Sellier1} we may roughly estimate $\xi _{1}\thickapprox 4$ $nm$
, $\xi _{2}\thickapprox 9$ $nm$, the magnetic scattering parameter $\alpha
\thickapprox 0.9$, $\xi _{f}\thickapprox 6$ $nm$ \ and $h\thickapprox 100$ $%
K.$ Extrapolating the expressions (\ref{I1}), (\ref{I2}) for low temperature
we have for the ratio $I_{2}/I_{1}\sim 0.1\exp (-d/\xi _{1})\sim 10^{-3}$
which \ is close to the observed value $3\cdot 10^{-3}\cite{Sellier2}$. On
the other hand, for the F-layer thickness $d\thickapprox 19$ $nm$ the $0$ - $%
\pi $ transition occurs at the temperature $5.3$ $K$ and the second harmonic
was too small to be observable. Smaller $\Delta /T$ ratio and larger F layer
thickness gives this case less favorable for the second harmonic observation.

We have demonstrated the presence of a small intrinsic second harmonic at
the $0$ $-\pi $ transition in S/F/S junction with a uniform barrier.
However, there is another mechanism of the negative second harmonic
generation due to the inhomogeneity of the F layer thickness \cite{Buzdin
Koshelev}. Indeed the roughness of F layer in the real S/F/S junctions \cite%
{Frolov}, \cite{Sellier1},\cite{Ryazanov2004} is of the order of $1$ $nm$.
This means that if the characteristic length $\delta l$ of the thickness
variation (along the contact surface) is larger than $d$, the critical
current will vary locally too. On the other hand, if $\delta l$ is much
smaller than the Josephson length $\lambda _{J}$, which for the current
density $10^{6}$ $A/m^{2}$ \cite{Ryazanov2004} is of the order of the
junction dimension ($50\times 50$ $\mu m^{2}$ in \cite{Ryazanov2004}), the
measured characteristics of the junction will be effectively averaged. At
the $0$ $-\pi $ transition we deal with a system where the local current
density is alternating $\pm I_{1}$ and $\overline{I_{1}}=0$. The resulting
local phase variation leads to the appearance of the \underline{negative}
second harmonic in the averaged current-phase relation \cite{Mints1},\cite%
{Buzdin Koshelev} $I_{2}\sim -\left\vert I_{1}\right\vert \left( \delta
l/\lambda _{J}\right) ^{2}$, where $\lambda _{J}$ is the Josephson length
corresponding to the current density $I_{1}$. The $1$ $nm$ roughness of the
F layer in the experiments \cite{Sellier2},\cite{Sellier1} permits to
estimate for $d\thickapprox 17$ $nm$ the value $\left\vert I_{1}\right\vert
\sim 5\cdot 10^{6}$ $A/m^{2}$ and $\lambda _{J}\sim (10-100)\mu m$. At the
present time there is no information about the characteristic length $\delta
l$ of thickness variation in the studied S/F/S junctions. Taking it as $1$ $%
\mu m$ for the $10\times 10$ $\mu m^{2}$ junction \cite{Sellier1},\cite%
{Sellier2} we have $I_{2}\sim -5(10^{2}-10^{4})$ $A/m^{2}$ while the
experimentally observed value is $\sim 3\cdot 10^{4}$ $\ A/m^{2}$ and the
sign of $j_{2}$ is unknown.

Let us now briefly discuss the thermodynamics of the $0$ $-\pi $ transition.
Near the transition temperature $T_{\pi }$, the amplitude of the first
harmonic $I_{1}$ may be considered as a linear function of $T$, i.e. $%
I_{1}=\mu \left( T-T_{\pi }\right) $, while the second harmonic term as a
temperature independent. The phase dependent contribution to the free energy
of junction (\ref{EJ}) being $E_{J}(\varphi )=\frac{\Phi _{0}}{2\pi c}\left[
-\mu \left( T-T_{\pi }\right) \cos \varphi -\frac{I_{2}}{2}\cos 2\varphi %
\right] .$For $I_{2}>0$ the transition occurs to be I order and at $T>T_{\pi
}$ ($0$ phase) $\delta F_{0}=-\frac{\Phi _{0}}{2\pi c}\mu \left( T-T_{\pi
}\right) $, while at $T<T_{\pi }$, in the $\pi $ phase\ $\delta F_{\pi }=-%
\frac{\Phi _{0}}{2\pi c}\mu \left( T_{\pi }-T\right) .$ Therefore the latent
heat of the transition is $q=\frac{\Phi _{0}}{\pi c}\mu T_{\pi }$. Taking
the parameters of the S/F/S junctions \cite{Sellier2},\cite{Sellier1}, we
may estimate $\mu T_{\pi }\sim 10^{-4}$ $A$ and then $q\sim 3\cdot 10^{-20}$ 
$J.$ The S/F/S junctions studied in \cite{Ryazanov2005} with the $0$ - $\pi $
transition at $d\thickapprox 11$ $nm$ reveal the parameter $\mu T_{\pi }$
and consequently latent heat, which must be two order of magnitude larger.

In the case $I_{2}<0$ the $0$ - $\pi $ transition is continuous and in the
interval $-\left| I_{2}\right| <I_{1}<\left| I_{2}\right| $ the equilibrium
phase difference is determined by $\cos \varphi =\mu \left( T-T_{\pi
}\right) /\left| I_{2}\right| $. The specific heat of this $\varphi -$ phase
is $\delta C=\frac{\Phi _{0}}{2\pi c}T\frac{\mu ^{2}}{2\left| I_{2}\right| }%
. $ Therefore we may expect on experiment an increase of the specific heat
by $\delta C$ in the narrow temperature region $T_{\pi }-\mu \left|
I_{2}\right| <T<$ $T_{\pi }+\mu \left| I_{2}\right| $. If we suppose that
the $0$ - $\pi $ transition observed in \cite{Sellier2} is continuous , we
may estimate $\delta C$ in the $\varphi -$ phase. Taking the experimental
value of $I_{2}\sim 3$ $\mu A$ we have $\delta C\sim 10^{-18}$ $J/K=$ $aJ/K$%
. The recent precise measurements of the specific heat of the
superconducting microrings \cite{Bourgeois} demonstrated the possibility to
register a specific heat variation of the order of $0.1$ $aJ/K$ per ring.

It may be easily verified that (independently on the sign of $I_{2}$ and
then on the scenario of the transition) the critical current varies near the
transition temperature ($\left| I_{1}\right| <<\left| I_{2}\right| $) as $%
I_{c}(T)=\left| I_{2}\right| \left( 1+\frac{\mu \left| T-T_{\pi }\right| }{%
\sqrt{2}\left| I_{2}\right| }\right) .$ The temperature dependence is linear
and the slope $dI_{c}/dT$ near $T_{\pi }$ is $1/\sqrt{2}$ time smaller than
that far way from $T_{\pi }$.

At $T=$ $T_{\pi }$ the current-phase relation is $j\left( \varphi \right)
=I_{2}\sin 2\varphi $ and then the AC Josephson effect would imply the
frequencies two time larger $\omega \rightarrow 2\omega =2\left( \frac{2e}{%
\hslash }V\right) .$ Note that this circumstance was responsible for the
observation of the half-integer Shapiro steps at $0$ - $\pi $ transition in
experiments \cite{Sellier2} and provided an additional proof of the nonzero
critical current at $T=$ $T_{\pi }$. The SQUID with such junctions would
have the periodicity $\Phi _{0}/2$ on the magnetic flux.

The structure of the soliton (Josephson vortex) is rather peculiar for the
long (along $y-$axis) junction near a $0$ - $\pi $ transition. For $I_{2}>0$%
, the phase distribution is determined by the following equation

\begin{equation}
\frac{d^{2}\varphi }{dy^{2}}=\frac{1}{\lambda _{J0}^{2}}\left( \varepsilon
\sin \varphi +\sin 2\varphi \right) ,
\end{equation}
where $\lambda _{J0}^{-2}=c\Phi _{0}S/\left( 8\pi ^{2}t\left\vert
I_{2}\right\vert \right) $, $\varepsilon =I_{1}/I_{2}$, and $t$ is the
effective junction thickness. The soliton-type solution is

\begin{equation}
\varphi =\arccos \left( 1-\frac{2\left( 1+\varepsilon \right) }{%
1+\varepsilon \cosh ^{2}(\sqrt{1+\varepsilon }y/\lambda _{J0})}\right) .
\end{equation}
The variation of the shape of the soliton is presented in Fig. 2.
Approaching the transition the central part of the soliton with $\varphi
\approx \pi $ grows and finally at $I_{1}=0$ the system has two degenerate
ground states $\varphi =0$ and $\varphi =\pi $. If $I_{2}<0$, the $\varphi - 
$junction is realized in the interval $-\left\vert I_{2}\right\vert
<I_{1}<\left\vert I_{2}\right\vert $. The S/F junctions near the temperature 
$T=$ $T_{\pi }$ could provide an excellent possibility to study the unusual
properties \cite{Buzdin Koshelev},\cite{Mints1} of these junctions.

In summary, we present an analytical solution of the problem of the second
harmonic contribution to the current-phase relation of the S/F/S JJs in
diffusive limit in the presence of uniaxial magnetic scattering. Note that
very recently the case of the isotropic magnetic scattering have been
studied numerically in \cite{Houzet} with qualitatively similar results. An
important conclusion of our work is that the $0$ - $\pi $ transition is
discontinuous for the S/F/S JJs with a homogeneous F barrier but may be
continuous in real junctions with modulated F layer thickness. In the latter
case a very special $\varphi -$junction exists in the transition region. The
modern microcalorimetric technique could be used for the experimental study
of the thermodynamics of the $0$ - $\pi $ transition and determines its type.

The author thanks M. Houzet for valuable comments and helpful suggestions.
Useful discussions with M. Kupriyanov, O. Bourgeois, M.\ Faure, E.Goldobin,
M. Kulic, Ch. Meyers, V. Ryazanov, and D.Van Harlingen are also gratefully
acknowledged.

\end{document}